\documentclass[epjCONF]{svjour}
\usepackage{graphicx}
\usepackage[varg]{txfonts} 
\usepackage[latin1]{inputenc}
\session-title{%
19$^{\textnormal{\footnotesize th}}$ International %
IUPAP Conference on Few-Body Problems in Physics%
}
\newcommand{\Tr}{{\rm Tr}}
\newcommand{\half}{\textstyle\frac{1}{2}}
\newcommand{\nn}{\nonumber}
\begin{document}
\title{%
Neutron-Proton Mass Difference in Nuclear Matter and in Finite Nuclei and the Nolen-Schiffer Anomaly
}%
\author{%
U.-G. Mei{\ss}ner\inst{1,2,3}
\and 
A.M. Rakhimov\inst{4,5}
\and
A. Wirzba\inst{2,3}\fnmsep\thanks{\email{a.wirzba@fz-juelich.de}}
\and
U.T. Yakhshiev\inst{2,6} 
}
\institute{%
Helmholtz-Institut f\"{u}r Strahlen- und Kernphysik
and Bethe Center for Theoretical Physics,
Universit\"{a}t Bonn, D-53115 Bonn, Germany
\and
Institut f\"{u}r Kernphysik and J\"{u}lich Center for Hadron Physics,
Forschungszentrum J\"{u}lich, D-52425 J\"{u}lich,  Germany
\and
Institute for Advanced Simulation, Forschungszentrum J\"{u}lich, D-52425 J\"{u}lich, 
Germany 
\and
Institute of Nuclear Physics, Academy of Sciences of Uzbekistan, Tashkent-132, Uzbekistan
\and
Institute of Physics and Applied Physics, Yonsei University, Seoul, 120-749, Korea
\and
Physics Department and Institute of Applied Physics, National University of Uzbekistan, Tashkent-174, Uzbekistan
}
\abstract{
The neutron-proton mass difference in (isospin asymmetric) nuclear matter and finite nuclei is studied in the framework of a medium-modified Skyrme model. 
The proposed effective Lagrangian incorporates both the medium influence of the surrounding nuclear environment on the single nucleon properties and an explicit isospin-breaking effect in the mesonic sector. Energy-dependent charged and neutral pion optical potentials in the s- and p-wave channels are included as well.
The present approach predicts that the neutron-proton mass difference is mainly dictated by its strong part and that it markedly decreases in neutron matter.  
Furthermore, the possible interplay between the effective nucleon mass in finite nuclei and the Nolen-Schiffer anomaly is discussed. In particular, we find that a correct description of the properties of mirror nuclei leads to a stringent restriction of possible modifications of the nucleon's effective mass in nuclei.
} 
\maketitle
%
%
%
\section{Introduction}
\label{WirzbaA_sec:intro}

This presentation is about
a calculation of the neutron-proton mass difference
in finite nuclei and in nuclear matter
in the framework of a medium-modified Skyrme model.

The nuclear-density dependence of isospin-breaking effects belongs to one of the fundamental
questions in nuclear physics~\cite{WirzbaA_Li97,WirzbaA_Ba05,WirzbaA_Stei04}. 
In particular, 
the effective neutron-pro\-ton mass difference in finite nuclei and nuclear matter 
$\Delta m_{\rm np}^*$,  although it is of astrophysical relevance, of importance for the
description of mirror nuclei, the stability of drip-line nuclei etc., is not very well under control.
Rather, there exist very different theoretical predictions of this quantity for isospin-asymmetric
nuclear mater
\cite{WirzbaA_LQ88,WirzbaA_Bo91,WirzbaA_Cha97,WirzbaA_Ku97,WirzbaA_Zu01,WirzbaA_Ts99,WirzbaA_Gr00,WirzbaA_Hof0,WirzbaA_Liu1,WirzbaA_Zu05,WirzbaA_vD05,WirzbaA_Le06,WirzbaA_vD06,WirzbaA_Ch07} 
which predict both\\  
qualitatively and quantitatively different results.

Most notably, the in-medium neutron-proton mass difference is relevant for the description of mirror nuclei and the  Nolen-Schiffer 
anomaly (NSA)~\cite{WirzbaA_Nolen,WirzbaA_Shlomo}.
Although there
are many theoretical approaches devoted to the explanation of the NSA discrepancy~\cite{WirzbaA_Sh82,WirzbaA_He89,WirzbaA_Ha90,WirzbaA_Wi85,WirzbaA_Co90,WirzbaA_Sh94,WirzbaA_Horo00,WirzbaA_Suz92,WirzbaA_Sch93,WirzbaA_Dr94,WirzbaA_Ad91,WirzbaA_Agr01}
this phenomenon is still not fully understood.

Skyrme-soliton models have the inherent advantage, compared with other hadronic models,
that they are based on chiral input (chiral Lagrangians in the meson sector) and that
they treat the {\em structure}, {\em properties} and {\em interactions} of the investigated 
nucleons on an equal footing\\ 
\cite{WirzbaA_Skyrme1,WirzbaA_Skyrme2,WirzbaA_Adkins1,WirzbaA_Adkins2,WirzbaA_Zahed,WirzbaA_MeiZahed}. 
Moreover, Skyrme-soliton models  allows for an easy incorporation of the nuclear background
via local medium-dependent coefficients~\cite{WirzbaA_Ulug2001,WirzbaA_Ulug2002,WirzbaA_Ulug2004}.
They are therefore well sui\-ted to address the problem of the me\-dium-dependence of
the effective neutron-proton mass splitting and the Nolen-Schiffer anomaly.
In Refs.~\cite{WirzbaA_us1,WirzbaA_us2} we investigated  $\Delta m_{\rm np}^*$ in
symmetric and asymmetric nuclear matter, respectively. Here we rather focus on the
results of Ref.~\cite{WirzbaA_us3} which addresses  the behavior of this
quantity in finite nuclei and the
Nolen-Schiffer anomaly.

The presentation is organized as follows. In Sect.~\ref{WirzbaA_sec:ModSkyrme} we review
how the Skyrme model should be generalized in order to include explicit isospin breaking and
explicit chiral breaking on the same footing. An unconventional breaking mechanism is introduced
and the relevant quantization procedure is discussed.
Section~\ref{WirzbaA_sec:Medium} is dedicated to the introduction of medium-modifications
to the isospin-extended Skyrme model. We review the construction of 
the medium-modified Skyrme
model based on  the pion dispersion in a nuclear background.
Section~\ref{WirzbaA_sec:Center} focusses on
the static results of the nucleon when the pertinent skyrmion is placed at the center of spherical
symmetric nuclear core. The construction of the electromagnetic part of the in-medium neutron-proton mass splitting is indicated.
Values for the effective proton mass, the (total and electromagnetic
contribution of the) in-medium neutron-proton
mass splitting, the proton and neutron magnetic moments and the scalar and isovector
rms-radii are reported.  
 In Sect.~\ref{WirzbaA_sec:Off-center} we discuss the in-medium
neutron-proton mass difference when the nucleon is located at a distance $R$ from the
center.  Corresponding results in infinite nuclear matter can be found in
Sect.~\ref{WirzbaA_sec:NM}.
 In Sect.~\ref{WirzbaA_sec:tentative} tentative conclusions are given to 
the behavior of the effective shift $\Delta m_{\rm np}^*$. The Nolen-Schiffer anomaly
and the prediction of the medium-modified  generalized Skyrme model are discussed
in Sect.~\ref{WirzbaA_sec:NSA}. We end the presentation with final remarks in Sect.~\ref{WirzbaA_sec:final}.

\section{Isospin-modified Skyrme model} \label{WirzbaA_sec:ModSkyrme}

\subsection{Generalized symmetry breaking}

The  standard Skyrme model\,\cite{WirzbaA_Skyrme1,WirzbaA_Skyrme2,WirzbaA_Adkins1,WirzbaA_Adkins2,WirzbaA_Zahed,WirzbaA_MeiZahed} 
consists of the  non-linear $\sigma$-model Lagrangian, which is of 
second order in the derivatives, and a fourth order soliton-stabilizing term: 
\begin{eqnarray}
   {\cal L}&=&\frac{F_\pi^2}{16}\Tr\left( \partial_\mu U \partial^\mu U^\dagger \right)
   +\frac{1}{32e^2}\Tr[U^\dagger \partial_\mu U,U^\dagger\partial_\nu U]^2 \!,
   \label{WirzbaA_eq:origSkyrme}
\end{eqnarray}   
where \begin{equation}
   U(x) = \exp\bigg(2 {\rm i} \sum_{a=1}^3 \tau^a \pi^a(x)/F_\pi\bigg) = 1 +{\rm i}\frac{2\tau\cdot\pi}{F_\pi} + \cdots
\end{equation}
is the usual chiral SU(2) matrix formulated in terms of the pion field $\pi^a$. The 
pion-decay constant $F_\pi$ (its physical  value would have been 186\,MeV) and  the stabilizing parameter $e$ are normally  -- and also here -- adjusted in such a way,
that,  after rigid-rotator quantization in the (baryon number) $B$=1 soliton sector, 
the empirical  isospin-averaged nucleon and delta masses are fitted.
Note that the usual Skyrme model 
preserves isospin symmetry. This holds even if  there is a chiral symmetry breaking term 
 \begin{equation}
  {\cal L}_{\rm SB} = \frac{F_\pi^2 m_\pi^2}{16} \Tr\left( U + U^\dagger -2\right)\,
 \end{equation}
present, which induces the pion-mass term in the ($B$=0) meson-sector of the model,
 see especially \cite{WirzbaA_Skyrme2,WirzbaA_Adkins2,WirzbaA_Zahed,WirzbaA_MeiZahed}.
In order to incorporate explicit chiral symmetry breaking and isospin-breaking on equal
footing as it is the case for the free pion Lagrangian
\begin{eqnarray*}
 {\cal L}_{\rm mes}&=& 
 \partial_\mu\pi^+\partial^\mu\pi^- -m_{\pi^\pm}^2\pi^+\pi^- + \half 
 \left(\partial_\mu\pi^0
 \partial^\mu\pi^0- m_{\pi^0}^2\pi^0 \pi^0\right),
 \\
 \pi^\pm&=&{\textstyle\frac{1}{\sqrt{2}}} (\pi_1\mp i\pi_2)\,,\quad
 \pi^0=\pi_3\,,
  \label{WirzbaA_eq:Lfree}
\end{eqnarray*}
the original Skyrme model has to be modified.
Following Rathske\,\cite{WirzbaA_Rathske} we therefore add to the ordinary Skyrme Lagrangian
(\ref{WirzbaA_eq:origSkyrme}) the following generalized symmetry breaking term 
\begin{eqnarray}
   {\cal L}_{{\rm g}\chi {\rm SB}}&=&-\frac{F_\pi^2}{16}\Bigl\{
   \Tr\left[(U-1){\overline{\cal M}}^2(U^\dagger-1)\right] \nn \\
  &&  -\Tr\left[(U-1)
   \tau_3{\cal M}_{\Delta}^2 (U^\dagger-1) {\tau_3}\right]\Bigr\},
  \label{WirzbaA_eq:genSB}
\end{eqnarray}
which explicitly breaks chiral symmetry and isospin symmetry.
Here the following definitions have been applied:
\begin{equation}
  {\overline{\cal M}}^2 \equiv \half ( m^2_{\pi^\pm}+ m^2_{\pi^0} ) \,,
   \quad {{\cal M}_{\Delta}^2 \equiv \half(m^2_{\pi^\pm}- m^2_{\pi^0})}
 \end{equation}  
with  $m_{\pi^\pm}\equiv m_{\pi^\pm}^{\rm strong}$ denoting the strong-interaction part of the mass of the charged pions. Note  that  the electromagnetic contribution to the latter mass is beyond the framework of the present model
and will  not  be considered here.

\subsection{Quantization} \label{WirzbaA_sec:Quantization}
Following Refs.~\cite{WirzbaA_us1,WirzbaA_us2,WirzbaA_us3}, 
we quantized the isospin-modified 
Skyrme model with the help of one constrained ($\varphi_3^c$(t)) and
three unconstrained time-dependent collective coordinates ($\varphi_1(t),\varphi_2(t),\varphi_3(t)$)
\begin{equation}
U \to A U A^\dagger \quad \mbox{with}\quad  
A\to A(t)=\exp\Big( {\rm i}\,\mathbf{\tau} \cdot\mathbf{\varphi}/2\Big)
\end{equation}
where  
\begin{equation}
{\dot\varphi_1=\omega_1,\quad \dot\varphi_2=\omega_2,\quad\dot\varphi_3=
\omega_3} +{\dot\varphi^c_3} \equiv {\omega_3} +{a_c} 
\end{equation}
are the  pertinent angular velocities.
Under the spherical hedgehog ansatz
$U(\vec r) = \exp\left ( {\rm i} \vec\tau\cdot \hat \vec r F(r)\right)$,
there exist also collective coordinates linked
to static rotations of the skyr\-mion in coordinate space, where
$\Omega_1, \Omega_2, \Omega_3$ are the  relevant angular velocities. 
In terms of both classes of angular velocities the spatially integrated
Skyrme Lagrangian (\ref{WirzbaA_eq:origSkyrme}) including the 
generalized symmetry breaking term (\ref{WirzbaA_eq:genSB})
turns into
\begin{eqnarray}
L&\equiv&  \int  {\rm d}^3 r \left\{  {\cal L}_2  + {\cal L}_4+ {\cal L}_{{\rm g}\chi{\rm SB}} \right\}
 = - M_{\rm NP} - {\cal M}_{\Delta}\Lambda_2 \nn\\
&&+   \half(\overbrace{\Lambda_2+\Lambda_4}^{\Lambda})\,
\Bigl[ \left( \omega_1\!-\!\Omega_1\right)^2 + \left(\omega_2\!-\!\Omega\right)^2 
+ \left(\omega_3\!-\!\Omega_3\!+\!a_c\right)^2\Bigr] \nn \\
&=&-M_{\rm NP}
+\half\Lambda \sum_{i=1}^3 \left( \omega_i-\Omega_i \right)^2 
+\Lambda \left(\omega_3 -\Omega_3\right)a_c,
 \label{WirzbaA_eq:intSky}
\end{eqnarray}
where $M_{\rm NP}$ is the static soliton mass  when it is non-perturbed (NP) by isospin-breaking. It is the
hedgehog mass in case the above-mentioned hedgehog ansatz is inserted for the time-independent chiral matrix.
$\Lambda_2$ and $\Lambda_4$ are the moments of inertia of the non-linear $\sigma$ model
Lagrangian ${\cal L}_2$  and the fourth-order stabilizing term ${\cal L}_4$, respectively.
In the last line of Eq.~(\ref{WirzbaA_eq:intSky}), we employed 
the constraint
\begin{equation}
  {\cal M}_{\Delta}\Lambda_2 = \half \Lambda a_c^{2}\,,
  \label{WirzbaA_eq:constraint}
\end{equation}
whose role is to balance the isospin-breaking terms, such that the hedgehog ansatz of
the usual Skyrme model can still be applied.
In terms of the  canonical conjugate momentum operators  $\hat T_i \equiv \partial L/\partial \omega_i$  (and $\hat J_i \equiv \partial L/\partial \Omega_i = -\hat T_i$) corresponding to isospin (and spin),
the
pertinent Hamiltonian becomes
\[
H =M_{\rm NP} + \half \Lambda a_c^{2} + {\frac{ {{\hat {\mathbf T}}^2}}{2\Lambda}} - {a_c \hat T_3}\,,
\]
while the baryon states are $|T, T_3, J=T, J_3= -T_3\rangle$.
When the Hamiltonian is sandwiched by these states, the energy of a nucleon of isospin component $\pm \half$ is determined as
\begin{equation}
m_{{\rm p}/{\rm n}} \equiv E_{{\rm p}/{\rm n}}=M_{\rm NP} + \half \Lambda a_c^{2} + {\frac{ 3}{8\Lambda}} \mp {a_c \half}\,.
\end{equation} 
The latter equation allows to isolate
the strong part of the neutron-proton mass difference as
$\Delta m_{\rm n p}^{\rm strong} = a_c$ with
\begin{equation}
a_c= \sqrt{ \frac{2{\cal M}_{\Delta}}{\Lambda}} \approx (1.291+0.686)\,{\rm MeV}\,.
\end{equation}
For consistency, the empirical electromagnetic contribution   $-0.686\,{\rm MeV}$ to
the in-vacuum value  $1.291$\,MeV of the
neutron-proton mass difference
had to be subtracted.

\section{Medium-modifications} \label{WirzbaA_sec:Medium}

The explicit isospin-breaking term is only one part contributing to the neutron-proton mass difference in a nucleus or nuclear matter. The other part comes from the isospin asymmetry
of the nuclear background. In order to include this  asymmetry into
the nuclear-background parameters of a medium-modified Skyrme model, used in
Refs.~\cite{WirzbaA_Ulug2001,WirzbaA_Ulug2002,WirzbaA_Ulug2004},
we start out by introducing
medium-modi\-fications to 
the free pion Lagrangian (\ref{WirzbaA_eq:Lfree}).

Note that throughout this presentation we will denote medium-modified Langrange terms,
energies, masses, moments of inertia, form factors, etc.\
with an explicit asterix.
\subsection{Pion dispersion in a nuclear background}
\label{WirzbaA_sec:Piondisp}

The pertinent medium-modified Lagrangian of the pion field can be formulated
in terms of the self-energies ({\it i.e.} energy-dependent optical potentials)
of the charged pions 
as follows:
\begin{eqnarray}
 {\cal L}_{\rm mes}^{*}&=&
  \sum_{\lambda=\pm,0}
  \Big\{\half\partial_\mu{\pi^\lambda}^\dagger
  \partial^\mu\pi^{\lambda}-\half{\pi^{\lambda}}^\dagger\Big(m_{\pi^\lambda}^2+
  {\hat\Pi^{{\lambda}}(\omega,\vec r) }\Big)\pi^\lambda\Big\} \nn
\\
  &=&{\cal L}_{\rm mes}-\half\bigg\{\pi_a {\frac{\hat\Pi^{-} + \hat\Pi^{+}}{2} }\pi_a
  +{\rm i}\, {\varepsilon_{ab3}}\,\pi_a\,
  {\frac{\hat\Pi^{-}- \hat\Pi^{+}}{2} }
  \,\pi_b\bigg\}.
  \label{WirzbaA_eq:Lmesstar}
\end{eqnarray}
In the local density approximation 
the $s$-wave self-energies 
become
\begin{equation}
   \hat\Pi^{\pm}_{s}(\omega,\vec r)= 
   -4\pi\,\Big(b_0^{\rm eff}(\omega)\,{\rho(\vec r)}\mp b_1(\omega)
   {\delta\rho(\vec r)}\Big)\, \eta 
\end{equation}   
with the total and isosvector density
\begin{equation}
  {\rho(\vec r)} = {\rho_n(\vec r)+\rho_p(\vec r)}\,,\quad
  {\delta\rho(\vec r)}= {\rho_n(\vec r)-\rho_p(\vec r)}
\end{equation}
and $\eta\equiv 1 + m_\pi/m_N$ in terms of the isospin-averaged pion and nucleon masses.
The effective isoscalar and isovector pion-nucleon scattering lengths
can be expressed as
\begin{eqnarray} 
  b_0^{\rm eff}(\omega)
&\approx& b_0(\omega)-\frac{3k_F}{2\pi}\big[b_0^2(\omega)+2b_1^2(\omega)\big]\,,
\\
b_0(\omega)&\approx&- \tilde b_0 \Bigl(1-m_\pi^{-2}\omega^2\Bigr)/(4\pi\eta)\,,
\\
b_1(\omega) &\approx& \tilde b_1
  \Bigl(m_\pi^{-1}\omega + 0.143 m_\pi^{-3}\omega^3\Bigr)/(4\pi\eta)
\end{eqnarray}
in terms of the total Fermi momentum $k_F$ and  the 
parameters  $\tilde b_0$ = $-1.206\,m_\pi^{-1}$, 
$\tilde b_1$ = $-1.115\,m_\pi^{-1}$. 
The corresponding $p$-wave self-energies read in the local density approximation
\begin{eqnarray}
  \hat\Pi^{\pm}_{p}(\omega,\mathbf{r})&=&
   \vec\nabla
   \frac{4\pi c^\pm(\omega,\mathbf{r})}{1+4\pi g^\prime c^\pm(\omega,\mathbf{r})}
  \cdot\vec\nabla \nn\\
 && \mbox{}
  -   \frac{4\pi\omega}{2m_N}\left(\vec\nabla^2c^\pm(\omega,\vec r)\right) 
\end{eqnarray}  
with $g'=0.47$ and 
\begin{equation} 
  c^\pm(\omega,\mathbf{r})\equiv \Big(c_0(\omega)\,{\rho(\vec r)}\mp 
  c_1(\omega)\,{\delta\rho(\vec r)}\Big)/\eta 
\end{equation}  
in terms of the isoscalar and isovector pion-nucleon scattering volumes
$c_0=0.21m\,_\pi^{-3}$ and $c_1=0.165\,m_\pi^{-3}$, respectively;
see Refs.\,\cite{WirzbaA_us2,WirzbaA_us3} for more details on the parameters and references.

\subsection{Medium-modified  generalized Skyrme model}
Marrying the generalized isospin-modified Skyrme model of  
Rathske~\cite{WirzbaA_Rathske} with the in-medium modified Skyrme mo\-del of
Refs.~\cite{WirzbaA_Ulug2001,WirzbaA_Ulug2002,WirzbaA_Ulug2004} (which is characterized
by density dependent coefficients of the second order symmetry-conser\-ving and breaking  Lagrangian terms) and extending it to asymmetric nuclear background as in Eq.~(\ref{WirzbaA_eq:Lmesstar}),
one arrives at the following isospin- and medium-modified Skyrme 
Lagran\-gian~\cite{WirzbaA_us1,WirzbaA_us2,WirzbaA_us3}:
\begin{eqnarray}
 {\cal L}^{{*}} &=& {\cal L}_2^{{*}}+{\cal L}_4                       
+{\cal L}_{\chi{\rm SB}}^{{*}} + 
\Delta{\cal L}_{\rm mes}+\Delta{\cal L}_{\rm env}^{*}\,,
 \label{WirzbaA_eq:genmedLag}
\\
{\cal L}_2^{{*}} &=& \frac{F_\pi^2}{16}
  \bigg\{
   \Big(1+\frac{{\chi_{s}^{02}}}{m_\pi^2}\Big)\,
   \Tr\left(\partial_0U\partial_0U^\dagger\right) \nn 
\\
  && \quad\mbox{}-\left(1-{\chi_{p}^0}\right)
   \Tr(\vec{\nabla} U\cdot\vec{\nabla} U^\dagger)\bigg\},
\nn
\\
  {\cal L}_4&=&\frac{1}{32e^2}\,\Tr\,[U^\dagger\partial_\mu U, U^\dagger \partial_\nu U]^2\,,
\nn
\\
  {\cal L}_{\chi {\rm SB}}^{{*}}&=&-\frac{F_\pi^2 m_{\pi}^2}{16}
  \big(1+m_\pi^{-2}{{\chi_{s}^{00}}}\big)
   \,\Tr\left[(U-1)(U^\dagger-1)\right],
\nn
\\
  \Delta{\cal L}_{{\rm mes}}&=&-\frac{F_\pi^2}{16}\sum_{a=1}^2 
  {\frac{m_{\pi^\pm}^2- m_{\pi^0}^2}{2}}
  \,\Tr(\tau_aU)\Tr(\tau_aU^\dagger),\quad\,\,
\nn
\\
  \Delta{\cal L}_{{\rm env}}^{{*}}&=&-\frac{F_\pi^2}{16}\sum_{a,b=1}^2
  {\varepsilon_{ab3}
  \frac{{\Delta\chi}_{s}+{\Delta\chi_{p}}}{2m_\pi}}\, \Tr(\tau_a U)\Tr(\tau_b\partial_0 U^\dagger)\,.
 \nn
\end{eqnarray}
Here ${\cal L}_{\rm mes}$ is the {\em explicitly} isospin-breaking Lagrangian, while 
$  \Delta{\cal L}_{{\rm env}}^{{*}}$ summarizes the
{\em enviroment-induced} isospin-breaking contribution. The second-order
Lagrangians ${\cal L}_2^*$ and
${\cal L}_{\chi {\rm SB}^*}$ are isospin-sym\-metric, but explicitly medium-dependent,
whereas ${\cal L}_4$ is neither isospin-breaking nor me\-dium-dependent. 
We preserve the original term for stabilizing skyrmions and do not modify it by, {\it e.g.} higher-order
derivative terms or vector-meson contributions, since it is  on the one hand generic for our purposes and on the other hand still simple to handle.

The medium functionals  appearing in ${\cal L}_2^*$ and ${\cal L^*}_{\!\!\chi{\rm SB}}$
are constructed by the rule  $\hat\Pi(\omega)\to\hat\Pi(i\partial_0)$  from the input of Sect.~\ref{WirzbaA_sec:Piondisp} and take the form:
\begin{eqnarray}
{\chi_{s}^{00}}&=&
\left(\tilde b_{0}+\frac{3k_F}{8\pi^2(1+m_\pi/m_N)}\tilde b_{0}^2\right){\rho}\,, 
\\
{\chi_{s}^{02}}  &=&
\left(\tilde b_0+\frac{3k_F}{4\pi^2(1+m_\pi/m_N)}\bigl(\tilde b_0^2-\tilde b_1^2\bigr)
\right){\rho},
\\
{\chi_p^0}&=&\frac{2\pi c^+}{1+4\pi g^\prime c^+}
+ \frac{2\pi c^-}{1+4\pi g^\prime c^-} 
\end{eqnarray}
with
\begin{eqnarray}
c^\pm&\equiv&\frac{c_0\rho\mp c_1\delta\rho} {1+m_\pi/m_N}\,,\\
{\Delta\chi_s}&=&\tilde b_1{\delta\rho}\,,\quad
{\Delta\chi_p}\ \, =\ \, -\frac{2\pi m_\pi}{m_\pi+ m_N}\, 
c_1\left(\vec\nabla^2{\delta\rho}\right) \,.
\label{WirzbaA_eq:gradient}
\end{eqnarray}
Note the explicit occurrence of gradient density terms in the last expression which
especially operate at the surface of finite nuclei and vanish in homogenous nuclear matter;
compare Ref.~\cite{WirzbaA_us3} with \cite{WirzbaA_us2}.
\subsection{Quantization of the in-medium Skyrme model}
By essentially copying the steps of Sect.~\ref{WirzbaA_sec:Quantization}, one can find the following expression
for the spatially integrated medium-dependent Lagrangian (\ref{WirzbaA_eq:genmedLag}), see
Refs.~\cite{WirzbaA_us2,WirzbaA_us3} for more details:
\begin{eqnarray*}
  \int {\rm d}^3r \,{\cal L}^*&=&-M_{\rm NP}^*-{{\cal M}_{\Delta}^2}\Lambda_{\rm mes}
  +\frac{\omega_1^2+\omega_2^2}{2}\Lambda_{\omega\omega,12}^{*}
\\
  &-&(\omega_1\Omega_1+\omega_2\Omega_2)\Lambda_{\omega\Omega,12}^{*}
  +\frac{\Omega_1^2+\Omega_2^2}{2}\Lambda_{\Omega\Omega,12}^{*}
\\
  &+&\frac{(\omega_3-\Omega_3+a^*)^2}{2}{\Lambda_{\omega\Omega,33}^{*}}
  +(\omega_3-\Omega_3+a^*){\Lambda_{\rm env}^{*}}\,.
\end{eqnarray*}
Here $\omega_i$  and  $\Omega_i$  are the angular velocities in 
isotopic and coordinate space, respectively, while  $\Lambda^*_{\omega\omega,i j}$,
$\Lambda^*_{\omega\Omega,i j}$ and $\Lambda^*_{\Omega\Omega,ij}$ are
the pertinent medium-dependent moments-of-inertia. Finally, $\Lambda^*_{\rm env}$ is
the medium-dependent coefficient of the angular velocity sum $(\omega_3 - \Omega_3 +a^*)$
generated by the environ\-ment-induced isospin breaking term $\Delta {\cal L}^*_{\rm env}$.

Constructing the pertinent Hamiltonian by standard means and sandwiching it
between the usual isospin-spin states,  
one can determined the energy of a nucleon (or delta) state classified by the total isospin T and third component
$T_3$ as
\begin{eqnarray}
E&=&M_{\rm NP}^*+{\cal M}_{\Delta}^2\Lambda_{\rm mes}
+\frac{\Lambda_{\rm env}^{*2}}{2\Lambda_{\omega\Omega,33}^*}\nn\\
&+&\frac{\Lambda_{\Omega\Omega,12}^*+\Lambda_{\omega\omega,12}^*
-2\Lambda_{\omega\Omega,12}^*}
{2(\Lambda_{\omega\omega,12}^*\Lambda_{\Omega\Omega,12}^*
-\Lambda_{\omega\Omega,12}^{*2})}\big(T(T+1)-T_3^2\big)\nn\\
&+&\frac{T_3^2}{2\Lambda_{\omega\Omega,33}^*}
-\left(a_c^*+\frac{\Lambda_{\rm env}^*}{\Lambda_{\omega\Omega,33}^*}\right) T_3\,.
  \label{WirzbaA_eq:Efin}
\end{eqnarray}
Consequently, the strong part of the 
neutron-proton mass difference in the interior of  a nucleus can be isolated from the
last term of Eq.~(\ref{WirzbaA_eq:Efin}) and takes the form
\begin{equation}
\Delta m_{\rm np}^{*(\rm strong)}
=a_c^*+\frac{\Lambda_{\rm env}^*}{\Lambda_{\omega\Omega,33}^*}
  \label{WirzbaA_eq:mnpstrong}
\end{equation}
with
${a_c^*} = \sqrt{2{\cal M}_{\Delta}^2\, {\Lambda_{\rm mes}} / {\Lambda^*_{\omega \Omega, 33}}}$.

\section{Skyrmion at the center of a nucleus} \label{WirzbaA_sec:Center}

The knowledge of Eqs.~(\ref{WirzbaA_eq:Efin}) and (\ref{WirzbaA_eq:mnpstrong}) allows to construct
the total mass of a proton (neutron) and the strong contribution to neutron-proton mass difference
when the nucleon is located at the core of a nucleus. 

In order to determined the total
in-medium neutron-proton mass splitting, the electric (E) and magnetic
(M) form factors of the nucleons must be utilized in addition:
\begin{eqnarray*}
G_{\rm E}^* (\vec q^2)&=&\half
\int {\rm d}^3r \, e^{i\vec q\cdot\vec r}j^0(\vec r)\,,\\
G_{\rm M}^* (\vec q^2) & = & \frac{m_{N}}{2}
\int {\rm d}^3r \, e^{i\vec q\cdot\vec r}[\vec r\times \vec j(\vec r)]\,,
\end{eqnarray*}
where $\vec q$ is the transferred momentum. Furthermore, $j^0$ and $\vec j$ correspond to the
time and space components of the properly normalized sum of the density-dependent 
baryonic current $B_\mu^*$ and the third component of the isovector current $\vec V^*_\mu$ of
the Skyrme model; for more details see Refs.\,\cite{WirzbaA_us2,WirzbaA_us3}. 

\begin{table*}[!hbt]
\caption{Static properties of a nucleon when the pertinent skyrmion is either in free space
of added to the center of a finite nucleus core, namely $^{14}$N, $^{16}$O, 
$^{38}$K, and $^{40}$Ca, respectively, such that a proton or  the total nucleus
$^{15}$O, $^{17}$F, 
$^{39}$Ca, and $^{41}$Sc results. 
  Here $m^*_{\rm p}$ is the in-medium proton mass, 
$\Delta m_{\rm np}^*$ is the in-medium neutron-proton mass difference and 
$\Delta m_{\rm np}^{*\rm (EM)}$ is its electromagnetic part. Furthermore, $\mu^*_{\rm p}$
and  $\mu^*_{\rm n}$ are the in-medium proton and neutron magnetic moments in units
of the free-space Bohr magnetons (n.m.). Finally,
 $\langle r^2\rangle^{*1/2}_{\rm E,S}$ and $\langle r^2\rangle^{*1/2}_{\rm E,V}$ are the in-medium
 isoscalar (S) and isovector (V) charge radii of the nucleon
inside a nucleus. For more details see Ref.\,\cite{WirzbaA_us3}.}
\label{WirzbaA_tab:static}  
\centering
\begin{tabular}{lrrrrrrr}
\hline\noalign{\smallskip}
Core\,$\to$\,Elem. & $m_{\rm p}^*$ \ \ \ & 
$\Delta m_{\rm np}^*$ &$\Delta m_{\rm np}^{*\rm (EM)}$& 
$\mu_{\rm p}^*\  \  \  $ & $\mu_{\rm n}^*$\  \  \   &
$\langle r^2\rangle^{*1/2}_{\rm E,S}$ & $\langle r^2\rangle^{*1/2}_{\rm E,V}$\\[0.5mm]
  & [MeV] & [MeV] & [MeV] & [n.m.] & [n.m.] & [fm]\ \ \ & [fm]\  \   \ \\
\noalign{\smallskip}
\hline
\noalign{\smallskip}
Vac.$\to $\,p &
938.268&1.291&-0.686&1.963&-1.236&0.481&0.739\\[2mm] 
$^{14}$N $\to$  $^{15}$O   &
593.285&1.668&-0.526&2.355&-1.276&0.656&0.850\\[0.5mm]
$^{16}$O $\to$ $^{17}$F   &
585.487&1.697&-0.517&2.393&-1.297&0.667&0.863\\[0.5mm]
$^{38}$K  $\to$ $^{39}$Ca  &
558.088&1.804&-0.480&2.584&-1.422&0.722&0.942\\[0.5mm]
$^{40}$Ca$\to ^{41}$Sc &
557.621&1.804&-0.478&2.569&-1.428&0.724&0.947\\
\noalign{\smallskip}
\hline
\end{tabular}
\end{table*}

The electromagnetic part of the neutron-proton mass difference is then calculated 
in terms of the scalar (S) and isovector (V)  electric (E)  and magnetic (M)  form factors according
to the formula 
\begin{eqnarray}
\Delta m_{\rm np}^{*\rm (EM)}&=& 
-\frac{4\alpha}{\pi}\int\limits_0^\infty {\rm d}q
\Big\{G_{\rm E}^{\rm S*}(\vec q\,^2)G_{\rm E}^{\rm V*}(\vec q\,^2)\nn\\
&&\mbox{}-\frac{\vec q\,^2}{2m_{N}^2}
G_{\rm M}^{\rm S*}(\vec q\,^2)G_{\rm M}^{\rm V*}(\vec q\,^2)\Big\}
\end{eqnarray}
of Ref.\,\cite{WirzbaA_Gasser}, where $\alpha$ is the electric fine structure constant.

Explicit expressions of the pertinent scalar and isovector
 charge densities ($\rho_{\rm E}^{\rm S,V}$) and magnetic densities
($\rho_{\rm M}^{S,V}$)  can be found in Ref.\,\cite{WirzbaA_us3}. 
The rms radii
$\langle r^2\rangle^{*1/2}_{\rm E,S}$ and $\langle r^2\rangle^{*1/2}_{\rm E,V}$ are calculated
from the charge densities in the standard way, whereas the in-medium 
magnetic moments of the proton and neutron
follow from calculating the sum and difference of the integrated scalar and isovector 
magnetic densities, respectively.
Explicit expressions for the in-medium skyrmion mass $M^*_{\rm NP}$ and the moments-of-inertia
appearing in Eqs.~(\ref{WirzbaA_eq:Efin}) and (\ref{WirzbaA_eq:mnpstrong}) can be found 
in Refs.~\cite{WirzbaA_us3}
as well.

In summary, the (in-medium) mass of the proton, the total and electromagnetic contribution
of the (in-medium) neutron-proton mass splitting, the (in-medium) magnetic moments of
the proton and neutron and the isocalar and isovector (in-medium) rms radii are listed
in Table~\ref{WirzbaA_tab:static} for the following situation:  either a skyrmion in free space 
or located at the center of a finite nucleus core ({\it e.g.}  $^{14}$N, $^{16}$O, $^{38}$N, $^{40}$Ca) is 
quantized as a proton (or neutron). In the case that it is quantized as a  proton, the resulting total nuclei
will be  $^{15}$O , $^{17}$F, $^{39}$Ca, $^{41}$Sc, respectively; in the case of a neutron,
the resulting nuclei are $^{15}$N, $^{17}$O, $^{39}$N, $^{41}$Ca instead.

In accordance with previous calculations for infinite nuclear matter
in the isospin-symmetric case~\cite{WirzbaA_us1},  the second and third columns of Table~\ref{WirzbaA_tab:static} show that 
the total and electromagnetic part, respectively, of the neu\-tron-proton mass difference are slightly increased
in finite nuclei relative to the vacuum case.  Furthermore, the
present model predicts  (see the first column of Table~\ref{WirzbaA_tab:static}) that 
the effective nucleon mass is strongly reduced at the center of the nucleus.
We will return to this point later on when we discuss the Nolen-Schiffer anomaly in
Sect.~\ref{WirzbaA_sec:NSA}.

Finally, the increase in the tabulated  in-medium 
values of the magnetic moments
and rms charge  radii is compatible with the hypothesis of the nucleon-swelling 
in a nuclear background which was already confirmed by  the isospin-sym\-metric
calculations of Refs.\,\cite{WirzbaA_Ulug2001,WirzbaA_Ulug2002,WirzbaA_Ulug2004} for the
me\-dium-modified Skyrme model; see also the static in-me\-dium results
of Refs.~\cite{WirzbaA_us1,WirzbaA_us2}. 

\section{Skyrmion  off-centered in a nucleus}   \label{WirzbaA_sec:Off-center}
\begin{figure}[!h]
\caption{A sketch of a skyrmion located inside a finite nucleus with 
$R=|\vec R|$ the separation between the geometrical center (O') of the skyrmion and
the center (O) of the nucleus. The vectors (angles) $\vec r$ ($\theta$) and $\vec r' $ ($\theta'$)
refer to the body-fixed coordinates of the nucleus and skyrmion, respectively. Since 
the nucleus is spherically symmetric, both coordinate systems can be orientated in such a way that
their $z$-axes coincide.}
\label{WirzbaA_fig:geo}       
 \centering
 \includegraphics[width=0.75\columnwidth]{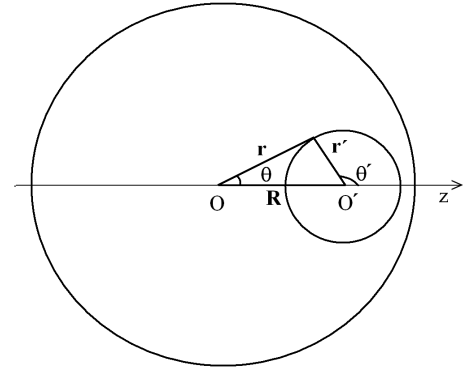}
\end{figure}

Whereas the results of the last section hold for the case that the skyrmion is 
at the center of a symmetrical core of a nucleus, the results of this section apply for the
more general situation that the skyrmion is located at a distance $R$ from the center of
a finite nucleus,  as specified in Fig.~\ref{WirzbaA_fig:geo}.
Note that in this case the spherical symmetric hedgehog ansatz cannot be used anymore,
since the background -- as viewed from the center of the {\em skyrmion} -- is not spherically symmetric. Thus there exists further deformations in
the isotopic and in the coordinate space~\cite{WirzbaA_Ulug2001,WirzbaA_Ulug2002}.
In this case a variational ansatz can still be used which allows for more freedom in
the radial and especially in the angular coordinates than the hedgehog form would do, but
is still compatible with the quantization procedure. See 
Ref.~\cite{WirzbaA_us3} for more details on this variational parameterization. In this reference,
it was also checked that the results of the variational computation merge with the ones
of the hedgehog computation when the skyrmion is moved back to the center of the nucleus.

\subsection{Effective proton mass $m_p^\ast$ inside a nucleus}
\begin{figure}[!h]
\caption{The dependence of the effective mass of the proton $m_{\rm p}^*$ on the distance $R$
between the center of the skyrmion and the center of the nucleus. The solid curve
represents the case that a skyrmion -- quantized as a proton -- is added to a $^{14}$N nucleus,
giving  $^{15}O$ in total. The dot-dashed curve
refers to  the case  of a $^{16}$O  nucleus $\to$  $^{17}$F in total.
The dashed curve stands for the case of $^{38}$K $\to$  $^{39}$Ca and the dotted
curve represents the case of $^{40}$Ca $\to$ $^{41}$Sc. 
The horizontal line marks the free space value of the proton
mass. 
Figure from Ref.~\cite{WirzbaA_us3}.}
\label{WirzbaA_fig:effmass}       

\bigskip
\centering
\includegraphics[width=0.75\columnwidth]{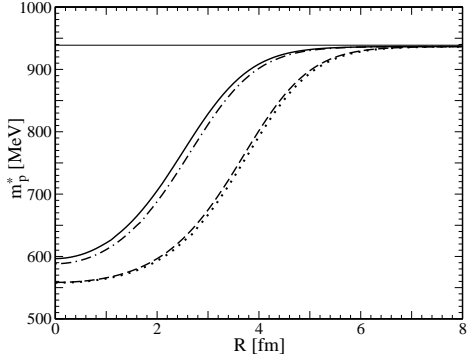} 
\end{figure}

First, we will show  in Fig.~\ref{WirzbaA_fig:effmass}
the results of Ref.~\cite{WirzbaA_us3} for the effective proton mass
$m_{\rm p}^*$ as function of the distances between the geometrical center of the pertinent
skyrmion and the center of the core-nucleus, {\it i.e.} 
$^{14}$N, $^{16}$O, $^{38}$K, and $^{40}$Ca. 
When the skyrmion is quantized as a proton, the resulting nuclei are
$^{15}$O, $^{17}$F, $^{39}$Ca, and $^{41}$Sc, respectively.
Note that the in-medium mass of the proton $m_{\rm p}^*$ starts with the value listed in
Table~\ref{WirzbaA_tab:static} when its is near the center of the nucleus -- in agreement
with the statements at the end of the last section.
With increasing
distance from the center of the nucleus the value of the effective mass  monotonically
increases, until it smoothly
approaches the free space value $m_{\rm p}$ at the border of the nucleus.

We will come back to the effective nucleon mass when we discuss the Nolen-Schiffer anomaly
in Sect.~\ref{WirzbaA_sec:NSA}. In the following we will concentrate on the effective
in-medium neutron-proton mass {\em difference}.

\subsection{Effective neutron-proton mass difference} 

\begin{figure}[!h]
\caption{The dependence of $\Delta m_{\rm np}^{* ({\rm strong)}}$ on the distance $R$ between
the center of the skyrmion and the center of the nucleus. The notations and input are the same
as in Fig.~\ref{WirzbaA_fig:effmass}, with the exception  that the horizontal line  marks
the free space value $\Delta m_{\rm np}^{({\rm strong})} = 2.0\, {\rm MeV}$. 
Figure from Ref.~\cite{WirzbaA_us3}.}
\label{WirzbaA_fig:strong} 

\bigskip
\centering
\includegraphics[width=0.75\columnwidth]{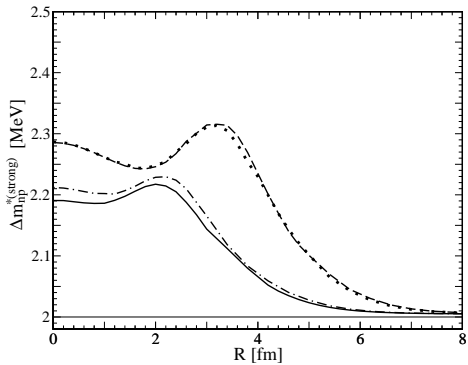}
\end{figure}

In Fig.~\ref{WirzbaA_fig:strong} the behavior of the strong part of the in-medium neutron-proton
mass difference can be found for the same notations and input as  in 
Fig.~\ref{WirzbaA_fig:effmass}.

Note that the strong part of the in-medium mass difference has a non-monotonic behavior.
This  follows from the fact that the density is a local quantity and that additional
isospin-breaking contributions arise due to the density gradients resulting from the
$p$-wave pion-nucleus scattering, see Eq.~(\ref{WirzbaA_eq:gradient}).  Especially, it can
be observed that in
the surface region of each nucleus, where the density gradients are large and the local isospin
asymmetry in the nuclear background is high, the value of $\Delta m_{\rm np}^{* ({\rm strong)}}$ 
is at an extremum.

\begin{figure}[!h]
\caption{The dependence of $\Delta m_{\rm np}^{* ({\rm EM)}}$ on the distance $R$ between
the center of the skyrmion and the center of the nucleus. The notations and input are the same
as in Fig.~\ref{WirzbaA_fig:effmass},
with the exception  that the horizontal line  marks
the free space value $\Delta m_{\rm np}^{({\rm EM})} = -0.69\, {\rm MeV}$. 
Figure from Ref.~\cite{WirzbaA_us3}.}
\label{WirzbaA_fig:EM}

\bigskip
\centering
\includegraphics[width=0.75\columnwidth]{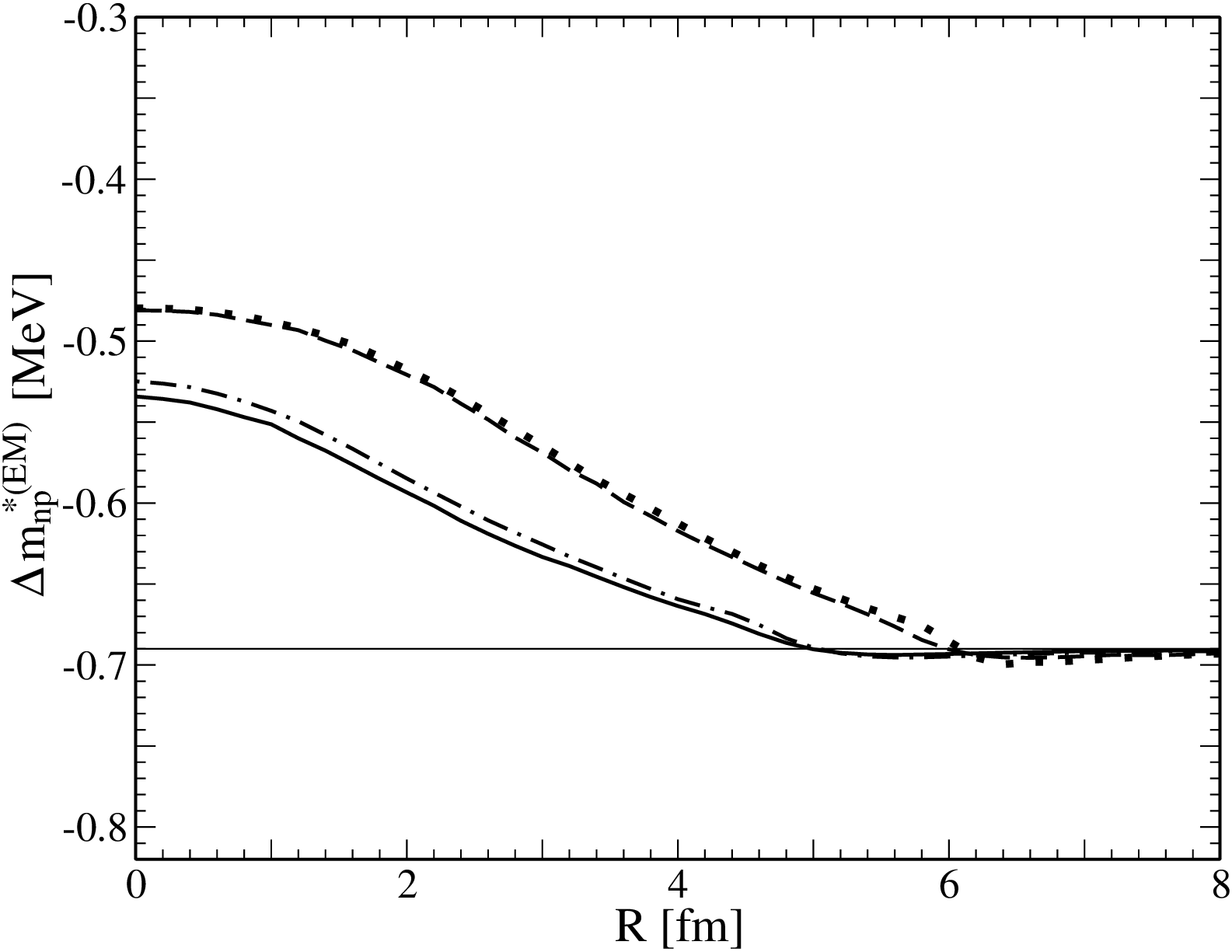} 
\end{figure}

In Fig.~\ref{WirzbaA_fig:EM} the electromagnetic part of the in-medium neutron-proton mass
difference is shown for the same nuclei as before. It can be seen that the variations in 
the electromagnetic part of the effective neutron-proton mass differences are small as
compared with their strong counter parts. Furthermore, $\Delta m_{\rm np}^{*({\rm EM})}$ nearly monotonically de\-creases with increasing distance from the center.
Note that the values at $R=0$ agree of course with the values listed in 
Table~\ref{WirzbaA_tab:static}.

\begin{figure}[!h]
\caption{The dependence of the total in-medium neutron-proton mass difference $\Delta m_{\rm np}^{*}$ on the distance $R$ between
the center of the skyrmion and the center of the nucleus. The notations and input are the same
as in Fig.~\ref{WirzbaA_fig:effmass}, with the exception  that the horizontal line  marks
the free space value $\Delta m_{\rm np} = 1.3\, {\rm MeV}$. 
Figure from Ref.~\cite{WirzbaA_us3}.}
\label{WirzbaA_fig:total}

\bigskip
\centering
\includegraphics[width=0.75\columnwidth]{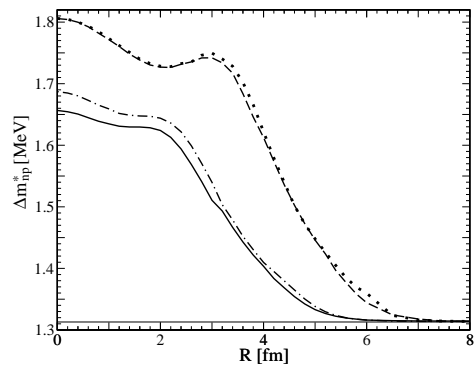}
\end{figure}

Finally, for completeness, in Fig.~\ref{WirzbaA_fig:total} the dependence of the total effective neutron-proton mass difference $\Delta m_{\rm np}^*$ on the distance $R$ is shown for the same nuclei as specified
above. Again,
the values at $R=0$ agree with the values listed in Table~\ref{WirzbaA_tab:static}.

\section{Neutron-proton mass difference in nuclear matter} \label{WirzbaA_sec:NM}

The densities of the finite nuclei discussed here and in Ref.~\cite{WirzbaA_us3} correspond
closely to the isosymmetric nuclear matter case studied in Refs.~\cite{WirzbaA_us1,WirzbaA_us2},
where a moderate increase of the effective neutron-proton difference, compatible with
Fig.~\ref{WirzbaA_fig:strong} or Fig.~\ref{WirzbaA_fig:total},
can be observed.
In neutron matter, however, as Ref.~\cite{WirzbaA_us2} shows, 
there is a pronounced decrease  of  the in-medium neutron-proton mass difference with increasing density. 

Therefore,
instead of discussing the den\-si\-ty-variations of all the static quantities in an infinite nuclear matter background as in Ref.~\cite{WirzbaA_us2}, we report  here only about the effective neutron-proton mass differences in nuclear matter studied in Ref.~\cite{WirzbaA_us2}.

\begin{figure}[!h]
\caption{Density dependence of the strong part $\Delta m_{\rm np}^{*(\rm
    strong)}$ of the neutron-proton mass difference in nuclear matter.  The abscissa represents
  the density $\rho$ normalized to the saturation density of ordinary nuclear
  matter $\rho_0=0.5m_{\pi}^3$, while the ordinate shows the mass difference
  in units of MeV.  The result in isospin-symmetric matter is plotted as a
  solid curve, the result of neutron-rich matter with $\delta\rho/\rho=0.2$ 
  as dashed curve, the dotted curve represents pure neutron matter
  ($\delta\rho/\rho=1$) and the dot-dashed curve proton-rich matter with
  $\delta\rho/\rho=-0.2$. Figure from Ref.~\cite{WirzbaA_us2}.}
\label{WirzbaA_fig:NM_strong}

\bigskip
\centering
\includegraphics[width=0.75\columnwidth]{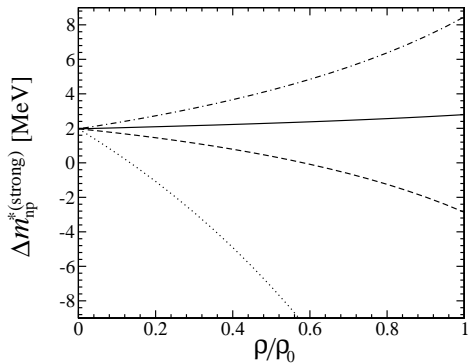}
\end{figure}

In Fig.~\ref{WirzbaA_fig:NM_strong} the strong part of the in-medium 
neutron-proton mass splitting in nuclear matter, $\Delta m_{\rm np}^*$,  is shown
for isospin-symmetric nuclear matter
(solid curve), neutron-rich matter (dashed curve), pure neutron matter
(dotted cur\-ve), and
proton-rich matter (dot-dashed curve). Especially, when the isospin symmetry of nuclear matter is broken,
 $\Delta m_{\rm np}^{*(\rm strong)}$ strongly varies    (see the dashed and dot-dashed curves in Fig.~\ref{WirzbaA_fig:NM_strong}).  
In pure neutron matter the change becomes very drastic 
(see the dotted curve in Fig.~\ref{WirzbaA_fig:NM_strong}), and 
$\Delta m_{\rm np}^{*(\rm strong)}$ 
decreases very rapidly with increasing density.

\begin{figure}[!h]
\caption{Density dependence of the electromagnetic part $\Delta m_{\rm
    np}^{*(\rm EM)}$ of the neutron-proton mass difference in nuclear matter. The axes and
  curves are defined as in Fig.~\ref{WirzbaA_fig:NM_strong}. 
  Figure from Ref.~\cite{WirzbaA_us2}.}
\label{WirzbaA_fig:NM_EM}

\bigskip
\centering
\includegraphics[width=0.75\columnwidth]{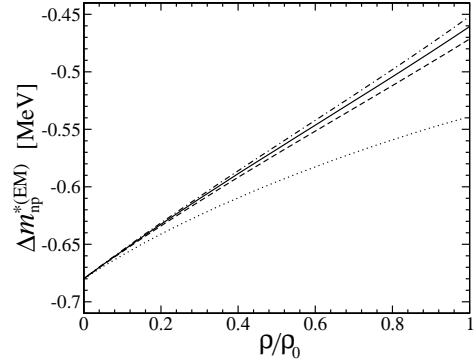}
\end{figure}

In  contrast to the strong part, 
the electromagnetic part of the neutron-proton mass 
difference  in nuclear matter varies only by a very small amount when the isospin-asym\-metry parameter $\delta \rho/\rho$
is increased (see Fig.~\ref{WirzbaA_fig:NM_EM}).

\begin{figure}[!h]
\caption{Density dependence of the  total
 neutron-proton mass difference   $\Delta m_{\rm np}^{*}$   in nuclear matter. The axes and
  curves are defined as in Fig.~\ref{WirzbaA_fig:NM_strong}. 
  Figure from Ref.~\cite{WirzbaA_us2}.}
\label{WirzbaA_fig:NM_total}

\bigskip
\centering
\includegraphics[width=0.75\columnwidth]{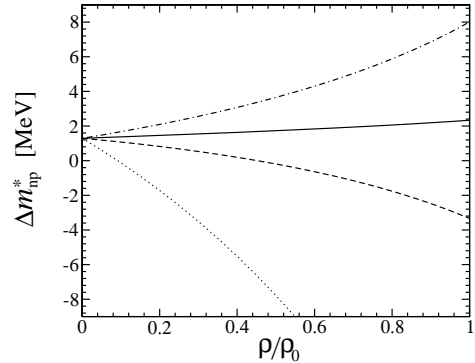}
\end{figure}

Again, for completeness, the total neutron-proton 
mass difference in nuclear matter is presented in Fig.~\ref{WirzbaA_fig:NM_total}. 
The difference to the purely strong case, shown in Fig.~\ref{WirzbaA_fig:NM_strong}, is hardly
visible.
\section{Tentative Conclusions}  \label{WirzbaA_sec:tentative}
In summary, we have  studied the effective neutron-proton mass
difference $\Delta m_{\rm np}^*$ in finite nuclei in the framework of
an isospin- and medium-modified Skyrme model. 

The in-medium mass of the proton starts near the center of the nucleus with 
a strongly reduced value and increases monotonically with increasing distance from the center. 
The strong part of the effective neutron-proton mass difference has a non-monotonic
behavior because of the additional isospin-breaking contributions 
due to the gradients arising
from the $p$-wave pion-nucleus scattering.
There is an extremum of $\Delta_{\rm np}^{*\,\rm strong}$ at the surface of the nucleus
because there the gradients are large and 
the isospin asymmetry in the nuclear background is high.
The electromagnetic part of the effective neutron-proton mass difference is negative and increases
monotonically in magnitude with increasing distance from the center. In magnitude its value
is small as compared with its strong counter part.
The other static quantities (in-medium magnetic moments, rms charge radii etc.) behave
according to the nucleon-swelling hypothesis for a nucleon embedded in the medium.

While in isosymmetric backgrounds the effective neu\-tron-proton mass difference
moderately increases, in asymmetric nuclear matter the strong part of this 
quantity can vary markedly (upwards in proton-rich matter and downwards in neutron-rich matter), whereas 
the electromagnetic contribution is always small and subleading.
Especially in neutron matter,   there is a strong decrease of
the  (strong and also the total) neutron-proton
mass difference with increasing density.

\section{Nolen-Schiffer anomaly} \label{WirzbaA_sec:NSA}

A long standing problem in nuclear physics is the Nolen-Schiffer anomaly  observed
in mirror nuclei~\cite{WirzbaA_Nolen,WirzbaA_Shlomo}.
Here we will show how the Nolen-Schiffer anomaly
can be treated in the framework of an isospin- and medium modified generalized Skyrme model.

The mass difference between mirror nuclei
\begin{equation}
  \Delta M\equiv{}^A_{Z+1}{M}_N-{}^A_{Z}{M}_{N+1}
\end{equation}
 which differ by one unit in their charges,  $\Delta Z =1$,
can very precisely be measured and 
is usually split into two terms
\begin{equation}
  \Delta M
 =\Delta E_{\rm EM}-\Delta m_{\rm np} 
  -\Bigl(\Delta m_{\rm np}^*-\Delta m_{\rm np}\Bigr)\,.
 \label{WirzbaA_eq:DeltaM}
\end{equation}
The first contribution is the
Coulomb energy difference $\Delta E_{\rm EM}$ computed relatively to the 
free neutron-proton mass difference, whereas the second is the in-medium neutron-proton mass
difference subtracted from the free one and therefore the net in-medium change of this quantity.
The Coulomb energy difference
includes various corrections, {\it e.g.}, by exchange terms, the center-of-mass motion, finite size
effects of the proton and neutron distributions, magnetic interactions, vacuum effects, short-range two-body correlations etc. It can be calculated with great accuracy (within 1\,\% error) 
\cite{WirzbaA_Shlomo}.  If $\Delta m_{\rm np}^*$ is assumed to be constant and equal to
the vacuum value, then Eq.~(\ref{WirzbaA_eq:DeltaM}) cannot be satisfied. This phenomenon
is called the Nolen-Schiffer-anomaly (NSA). Quantitatively, the NSA ranges -- throughout the
periodic table -- from
a few hundred keV for the lightest nuclei up to several MeV for the heaviest ones.
A possible resolution is the assumption that 
the effective neutron-proton mass difference would decrease with increasing mass number $A$,
such that
\begin{equation}
 \Delta_{NSA} = \Delta m_{\rm np} - \Delta m_{\rm np}^*\,.
\end{equation}

Within the present approach $\Delta m_{\rm np}^*$ has a local $R$ dependence 
according to the location of the nucleons inside the nuclei as shown
in Fig.~\ref{WirzbaA_fig:total}. 
In order to compare these results 
with the experimental data one therefore has to average the value of 
 $\Delta m_{\rm np}^*$ with respect to the separation $R$.
Since the nucleons inducing the Nolen-Schiffer anomaly are valence
nucleons, these must be located in the peripheral region of each of the
mirror nuclei, if the latter differ by one particle or hole from
a (magic) closed-shell nucleus.
The averaged effective masses and splittings 
 can therefore be expressed as follows:
 \begin{eqnarray}
  \overline{m}_{\rm n}^*&\equiv&\int 
     m_{\rm n}^*(R)\,\big|\psi_{\rm n}(R)\big|^2{\rm d}^3R\,, \nn \\
       \overline{m}_{\rm p}^*&\equiv&\int 
     m_{\rm p}^*(R)\,\big|\psi_{\rm p}(R)\big|^2{\rm d}^3R\ ,\nn \\
     \Delta \overline{m}_{\rm np}^* &\equiv& \overline{m}_{\rm n}^* - \overline{m}_{\rm p}^*\,,
 \label{WirzbaA_eq:DeltaMaver}
 \end{eqnarray} 
where $|\psi_{\rm n}(R)|^2$ 
and  $|\psi_{\rm p}(R)|^2$ are the density distributions of the in-medium 
neutron and proton, respectively.
In terms of the difference of the density distributions
\[
\Delta \psi_{\rm np}^2(R) \equiv    \bigl | \psi_{\rm n}(R) \bigr|^2
                                                           -\bigl | \psi_{\rm p}(R) \bigr|^2\,,
\]
Eq.~(\ref{WirzbaA_eq:DeltaMaver}) can be rewritten as
\begin{eqnarray}
  \Delta\overline{m}_{\rm np}^* &\approx&
 \int\! \!\left\{
   \Delta \psi_{\rm np}^2(R)\,
 m_{\rm p}^*(R) 
  +
  \Delta m_{\rm np}^*(R)\, | {\psi_{\rm p}}(R)|^2
  \right\}{\rm d}^3R \nn \\
  &\equiv& \Delta \overline{m}_{\rm np}^{*(1)}
                 +\Delta \overline{m}_{\rm np}^{*(2)}\,,
  \label{WirzbaA_eq:DeltaMtwo}
 \end{eqnarray} 
 where the subleading contribution of the cross term 
 \[
  \int \Delta \psi_{\rm np}^2 \,\Delta m_{\rm np}^*
 \,{\rm d}^3 R
 \]
is neglected. Thus here the Nolen-Schiffer anomaly simply reads
\begin{equation}
\overline\Delta_{\rm NSA} = \Delta m_{\rm np} - \Bigl( \Delta m_{\rm np}^{*(1)} 
+ \Delta m_{\rm np}^{*(2)} \Bigr ) \,.
\label{WirzbaA_eq:NSAdiscrepancy}
\end{equation}
It is listed in  Table~\ref{WirzbaA_tab:NSA}, left panel (labeled 
`$\alpha$=0') of the third column (labeled   `Present approach') for chosen pairs
of mirror nuclei, such that  the overall mass number $A$ increases.
\begin{table*}[!htb]
\caption{The averaged mass $\overline{m}_{\rm p}^*$
of the valence proton in a 
given nucleus,
the contributions to the effective neutron-proton mass difference 
(see Eq.~(\ref{WirzbaA_eq:DeltaMtwo})) and 
the  Nolen-Schiffer discrepancy  $\overline{\Delta}_{\rm NSA}$ calculated in the
present approach  by 
Eq.~(\ref{WirzbaA_eq:NSAdiscrepancy}) or Eq.~ (\ref{WirzbaA_eq:toy})
and the corresponding ``empirical'' results of Ref.~\cite{WirzbaA_Shlomo}.
All quantities are in units of MeV.
}
\label{WirzbaA_tab:NSA}
\centering
\begin{tabular}
{l | cc |  ccc | ccc | c l}
\hline\noalign{\smallskip}
\qquad&&&
\multicolumn{6}{|c|}{Present approach}& \\
\noalign{\smallskip}
\cline{4-9}
\noalign{\smallskip}
Nuclei&\multicolumn{2}{|c|}{$\overline{m}_{\rm p}^*$}
&\multicolumn{3}{c|}{{$\alpha_{\rm ren}=0$}}&
\multicolumn{3}{c|}{{$\alpha_{\rm ren}=0.95$}}&$\overline{\Delta}_{\rm NSA}^{\, \rm empirical}$\\
\noalign{\smallskip}
\cline{2-9}
\noalign{\smallskip}
&{$\alpha_{\rm ren}$=\,0}  & {$\alpha_{\rm ren}$=\,0.95}
&  $\Delta\overline{m}_{\rm np}^{*(1)}$  & $\Delta\overline{m}_{\rm np}^{*(2)}$ 
& $\overline{\Delta}_{\rm NSA}$ &
$\Delta\overline{m}_{\rm np}^{*(1)}$ &  $\Delta\overline{m}_{\rm np}^{*(2)}$ 
& $\overline{\Delta}_{\rm NSA}$
&Ref.~\cite{WirzbaA_Shlomo}\\
\noalign{\smallskip}
\hline
\noalign{\smallskip}
$^{15}$O-$^{15}$N&767.45&928.30&-4.27 & 1.56 &4.02&-0.21&1.33&0.20& $0.16\pm0.04$\\
$^{17}$F-$^{17}$O&812.35&930.54&-5.53 & 1.52 &5.33&-0.28&1.32&0.27 &$0.31\pm0.04$\\
$^{39}$Ca-$^{39}$K&724.78&926.16&-8.11& 1.67 &7.75&-0.41&1.33&0.37&  $0.22\pm0.08$\\
$^{41}$Sc-$^{41}$Ca&771.71&928.51&-9.74&1.62 &9.44&-0.49&1.33&0.47 & $0.59\pm0.08$\\
\noalign{\smallskip}
\hline
\end{tabular}
\end{table*}

Whereas qualitatively the calculated NSA values have the correct $A$ behavior, quantitatively the results are more than one order of magnitude too big (compare with the empirical
values listed in the last column of Table~\ref{WirzbaA_tab:NSA}).
This can be traced back to the pronounced  negative shift of $\Delta \overline m_{\rm np}^{*(1)}$
(see the first entry of the third column of Table~\ref{WirzbaA_tab:NSA}). This shift
is mainly there for three reasons:
\begin{description}
\item[(i)]
the rather large renormalization of the effective nucleon mass, 
\item[(ii)] 
the pronounced $R$ dependence of $m_{\rm p}^*$ inside the nucleus 
(see Fig.~\ref{WirzbaA_fig:effmass}), and
\item[(iii)]  the relative swelling of the proton distributions due to the 
Coulomb factor, {\em i.e.} $\Delta\psi^{2}_{\rm np}\ne 0$.
\end{description}  
For example, the averaged in-medium
mass of the valence proton in $^{17}$O is reduced to
$\overline{m}_{\rm  p}^*=812.35$~MeV. This drop of about  125\,MeV is
very large in comparison with the
empirical value of the binding energy per
nucleon in nuclear matter.    
For heavier nuclei, where the density in the interior approximates 
the normal nuclear matter density, 
the drop of the averaged effective mass is even
larger, {\em e.g.} $m_{\rm p}\!-\!\overline{m}_{\rm p}^*\sim$ (150 -- 200)\,MeV in
the $^{40}$Ca region (see the second column of Table~\ref{WirzbaA_tab:NSA}) down to
$\sim 300$~MeV in the $^{208}$Pb region.

If solely the contribution 
$\Delta \overline{m}_{\rm  np}^{*(2)}$ (due to the explicit $R$ 
dependence of the
neutron-proton mass difference) were considered, 
then the NSA in the present approach
would even have a  negative sign: 
$\Delta m_{\rm np}-\Delta \overline{m}_{\rm  np}^{*(2)}<0$.

Instead of driving the input parameters of Sect.~\ref{WirzbaA_sec:Piondisp} to unphysical values
in order to match the NSA discrepancy, 
we suggested in Ref. ~\cite{WirzbaA_us3}  to invert the problem and  to estimate
the effective nucleon mass inside finite nuclei according to the
results in the isospin-breaking sector.  To perform this task 
an artificially added  
renormalization parameter $\alpha_{\rm ren}$ in the expression 
\begin{equation}
 m_{\rm n,p}^*(R,\alpha_{\rm ren})=m_{\rm n,p}^*(R)
+\big(m_{\rm n,p}-m_{\rm n,p}^*(R)\big)\,\alpha_{\rm ren}
\label{WirzbaA_eq:toy}
\end{equation}
of the effective nucleon mass is fine-tuned in such a way, that
the Nolen-Schiffer anomaly is satisfied. The results are presented in
Table~\ref{WirzbaA_tab:NSA}, fourth column labeled   `$\alpha_{\rm ren}=0.95$'.  
It can be seen that
a successful  description of the correct order of the NSA implies 
a rather small drop of the
mass of the valence nucleons: 
$m_{\rm n,p}-\overline{m}_{\rm n,p}^*(\alpha_{\rm ren}=0.95)\sim 10$~MeV 
which is close to the
empirical binding energy per nucleon in nuclear matter.  
In this case, the
contribution to the Nolen-Schiffer anomaly from the term $\Delta \overline{m}_{\rm np}^{*(2)}$ 
can be neglected:
$ 
\Delta m_{\rm np}- \Delta\overline{m}_{\rm np}^{*(2)}(\alpha_{\rm ren}=0.95)\,
\sim -0.02\,{\rm MeV}$.

\section{Final remarks}  \label{WirzbaA_sec:final}
Let us discuss the relevance of our results for the Nolen-Schiffer
anomaly. Qualitatively, our approach predicts the sign and the relative mass-number increase of
this anomaly. But quantitatively it is far from satisfactory:
the results are more than one order of magnitude too large. Clearly, the part of our
calculation relevant to the Nolen-Schiffer anomaly depends on the
proton and neutron distributions of the mirror nuclei and is very
sensitive to the behavior of the wave functions of the valence
nucleons in the peripheral region of the nucleus. 
We have pointed out the possibility that
the Nolen-Schiffer anomaly may rather follow from
the behavior of the effective nucleon mass 
in finite nuclei than from the effective neutron-proton mass difference:
our calculations imply  that the Nolen-Schiffer anomaly 
could not and, maybe, should not be 
saturated by $\Delta \overline{m}_{\rm np}^{*(2)}$ (the averaged contribution due to the
explicit density and radial dependence of the neutron-proton mass difference).  
Rather more important is   
$\Delta \overline{m}_{\rm np}^{*(1)}$, the contribution 
due to the difference in the squared wave functions of valence
proton and neutron  weighted by  
the local  (density and density-gradient induced) 
variation of the effective mass of the nucleon. 
In fact, when we restrict the in-medium reduction
of the (averaged effective) proton 
mass to about 1\,\% of the free proton mass -- a value which is compatible with the empirical
binding energy per nucleon in nuclear matter -- 
we obtain a rather precise description of the NSA.
Here,  we should remark that the gradient
terms which are present in our model do not noticeably affect the
scaling behavior of $m^*$. They are important for the surface behavior
of $\Delta m_{\rm np}^{*}$, though.

The calculated 
Coulomb energy differences $\Delta E_{\rm EM}$ (\ref{WirzbaA_eq:DeltaM}) 
for mirror nuclei
of Ref.~\cite{WirzbaA_Shlomo} 
incorporate
a contribution due to the different wave functions
of valence nucleons that  is known as Thomas-Ehrman
 effect~\cite{WirzbaA_Thomas51,WirzbaA_Ehrman51}.  Note, however, that in 
 Ref.~\cite{WirzbaA_Shlomo}  a constant, $R$-independent value of the nucleon mass --
namely the free mass in vacuum --  was used, whereas here the effect is based on the $R$-{\em dependence}
of the effective nucleon mass. 

In summary,
the possibility exists that 
the anomaly  of the mirror nuclei  can be saturated by invoking a
dynamical (local) mass of the nucleon that needs to be only slightly reduced in 
comparison to its  free counter part.
In this context, it should be pointed out that the anomaly could also be removed by
the introduction of an additional term into the Coulomb part of the energy density
of nuclear systems in the 
local-density-functional approach \cite{WirzbaA_Bulgac96,WirzbaA_Fayans98,WirzbaA_Bulgac99,WirzbaA_Fayans01}.
In fact, this term is chosen such that
it is proportional to the  {\em isoscalar} rather than to the 
{\em isovector} density.  Furthermore, it should be stressed that the isoscalar 
contribution  is surface dominated. It is even 
argued in 
Refs.~\cite{WirzbaA_Brown98,WirzbaA_Brown00} that
the effective-to-free-nucleon-mass ratio $m_{N}^*/m_{N}$ is unity
to within a few percent. 
Apparently, different model calculations can lead to similar conclusions 
about the origin of the Nolen-Schiffer anomaly.

Returning to the in-medium modified Skyrme model, we would like to point out
that the results might still be improved if the calculations can be made more self-consistent,
{\it e.g.}, by the incorporation of
feedback mechanisms  between the modified skyrmion and the local nuclear
background. In addition,  the inclusion of further degrees of freedom might be 
a useful possibility as their introduction can anyhow be motivated
by more detailed considerations about the
nucleon structure and the nucleon-nucleon interaction.  Also the non-local character
of the effective  nucleon mass may be of importance.

\section*{Acknowledgement}
We would like to thank Horst Lenske for useful discussions and 
a computer code calculating 
nuclear densities of finite nuclei. We are also grateful to
Frank Gr\"ummer for providing us  with  
calculated shell-model wave functions.
The work of U.T.Y. was supported by the
Alexander von Humboldt Foundation.
The work of A.M.R. was supported by the second phase of
the Brain Korea 21 Project in 2007 and by the German Academic Exchange Service DAAD.
Partial financial support from the EU Integrated Infrastructure
Initiative Hadron Physics Project (contract number RII3-CT-2004-506078),
by the DFG (TR 16, ``Subnuclear Structure of Matter'') and by BMBF
(research grant 06BN411) is gratefully acknowledged.
This work is partially supported by the Helmholtz Association through
funds provided to the virtual institute ``Spin and strong QCD'' (VH-VI-231).



\begin{thebibliography}{99}

\bibitem{WirzbaA_Li97}
  B.A.~Li, C.M.~Ko and W.~Bauer,
  Int.\ J.\ Mod.\ Phys.\  E {\bf 7} (1998) 147
  [arXiv:nucl-th/9707014]

\bibitem{WirzbaA_Ba05}
  V.~Baran, M.~Colonna, V.~Greco and M.~Di Toro,
  Phys.\ Rept.\  {\bf 410}  (2005) 335
  [arXiv:nucl-th/0412060]

\bibitem{WirzbaA_Stei04}
  A.W.~Steiner, M.~Prakash, J.M.~Lattimer and P.J.~Ellis,
  Phys.\ Rept.\  {\bf 411} (2005) 325
  [arXiv:nucl-th/0410066]
  
\bibitem{WirzbaA_LQ88}
  M.~Lopez-Quelle, S.~Marcos, R.~Niembro, A.~Bouyssy and V.G.~Nguyen,
  Nucl.\ Phys.\  A {\bf 483} (1988) 479 

\bibitem{WirzbaA_Bo91}
  I.~Bombaci and U.~Lombardo,
  Phys.\ Rev.\  C {\bf 44} (1991) 1892


\bibitem{WirzbaA_Cha97}
  E.~Chabanat, J.~Meyer, P.~Bonche, R.~Schaeffer and P.~Haensel,
  Nucl.\ Phys.\  A {\bf 627}  (1997) 710


\bibitem{WirzbaA_Ku97}
  S.~Kubis and M.~Kutschera,
  Phys.\ Lett.\  B {\bf 399} (1997) 191
  [arXiv:astro-ph/9703049]

\bibitem{WirzbaA_Zu01}
  W.~Zuo, I.~Bombaci and U.~Lombardo,
  Phys.\ Rev.\  C {\bf 60} (1999) 024605
  [arXiv:nucl-th/0102035]

\bibitem{WirzbaA_Ts99}
  K.~Tsushima, K.~Saito and A.~W.~Thomas,
  Phys.\ Lett.\  B {\bf 465}  (1999) 36
  [arXiv:nucl-th/9907101]

\bibitem{WirzbaA_Gr00}
  V.~Greco, M.~Colonna, M.~Di Toro, G.~Fabbri and F.~Matera,
  Phys.\ Rev.\  C {\bf 64} (2001) 045203 (2001)

\bibitem{WirzbaA_Hof0}
  F.~Hofmann, C.M.~Keil and H.~Lenske,
  Phys.\ Rev.\  C {\bf 64} (2001) 034314 (2001)
  [arXiv:nucl-th/0007050]

\bibitem{WirzbaA_Liu1}
  B.~Liu, V.~Greco, V.~Baran, M.~Colonna and M.~Di Toro,
  Phys.\ Rev.\  C {\bf 65} (2002) 045201

\bibitem{WirzbaA_Zu05}
  W.~Zuo, L.G.~Cao, B.A.~Li, U.~Lombardo and C.W.~Shen,
  Phys.\ Rev.\  C {\bf 72} (2005) 014005 
  [arXiv:nucl-th/0506003]
  
\bibitem{WirzbaA_vD05}
  E.N.E.~van Dalen, C.~Fuchs and A.~Faessler,
  Phys.\ Rev.\ Lett.\  {\bf 95} (2005) 022302 
  [arXiv:nucl-th/0502064]

\bibitem{WirzbaA_Le06}
  T.~Lesinski, K.~Bennaceur, T.~Duguet and J.~Meyer,
  Phys.\ Rev.\  C {\bf 74} (2006) 044315 
  [arXiv:nucl-th/0607065]

\bibitem{WirzbaA_vD06}
  E.N.E.~van Dalen, C.~Fuchs and A.~Faessler,
  Eur.\ Phys.\ J.\  A {\bf 31} (2007) 29
  [arXiv:nucl-th/0612066]


\bibitem{WirzbaA_Ch07}
  L.W.~Chen, C.M.~Ko, B.A.~Li and G.C.~Yong,
  arXiv:0704.2340 [nucl-th]

  \bibitem{WirzbaA_Nolen}
  J.A.~Nolen and J.P.~Schiffer,
  Ann.\ Rev.\ Nucl.\ Part.\ Sci.\  \textbf{19} (1969) 471

\bibitem{WirzbaA_Shlomo}
  S.~Shlomo, Rep. Prog. Phys. {\bf 41} (1978) 957

\bibitem{WirzbaA_Sh82}
S.~Shlomo,
Physica Scr. {\bf 26} (1982) 280

\bibitem{WirzbaA_He89}
  E.M.~Henley and G.~Krein,
  Phys.\ Rev.\ Lett.\  {\bf 62} (1989) 2586 

\bibitem{WirzbaA_Ha90}
  T.~Hatsuda, H.~Hogaasen and M.~Prakash,
  Phys.\ Rev.\ Lett.\  {\bf 66} (1991) 2851
  [Erratum-ibid.\  {\bf 69} (1992) 1290]

\bibitem{WirzbaA_Wi85}
  A.G.~Williams and A.W.~Thomas,
  Phys.\ Rev.\  C {\bf 33}  (1986) 1070

\bibitem{WirzbaA_Co90}
  T.D.~Cohen, R.J.~Furnstahl and M.K.~Banerjee,
  Phys.\ Rev.\  C {\bf 43} (1991) 357

\bibitem{WirzbaA_Sh94}
  M.H.~Shahnas,
  Phys.\ Rev.\  C {\bf 50} (1994) 2346

\bibitem{WirzbaA_Horo00}
  C.J.~Horowitz, J.~Piekarewicz,
  Phys. \ Rev. \ C {\bf 63} (2000) 011303R

\bibitem{WirzbaA_Suz92}
  T.~Suzuki, H.~Sagawa and A.~Arima,
  Nucl. Phys. {\bf A536}  (1992) 141

\bibitem{WirzbaA_Sch93}
  T.~Schafer, V.~Koch and G.E.~Brown,
  Nucl.\ Phys.\  A {\bf 562} (1993) 644

\bibitem{WirzbaA_Dr94}
  E.G.~Drukarev and M.G.~Ryskin,
  Nucl.\ Phys.\  A {\bf 572} (1994) 560

\bibitem{WirzbaA_Ad91}
  C.~Adami and G.E.~Brown,
  Z.\ Phys.\  A {\bf 340} (1991) 93

\bibitem{WirzbaA_Agr01}
B.K.~Agrawal, T.~Sil, S.K.~Samaddar, J.N.~De and S.~Shlomo,
Phys.\ Rev.\ C {\bf 64}  (2001) 024305 

\bibitem{WirzbaA_Skyrme1}
  T.H.R.~Skyrme,
  Proc.\ Roy.\ Soc.\ Lond.\ A \textbf{260} (1961) 127

\bibitem{WirzbaA_Skyrme2}
  T.H.R.~Skyrme,
  Nucl.\ Phys.\  \textbf{31}  (1962) 556

\bibitem{WirzbaA_Adkins1}
  G.S.~Adkins, C.R.~Nappi and E.~Witten,
  Nucl.\ Phys.\ B \textbf{228}  (1983) 552

\bibitem{WirzbaA_Adkins2}
  G.S.~Adkins and C.R.~Nappi,
  Nucl.\ Phys.\ B \textbf{233}  (1984) 109

\bibitem{WirzbaA_Zahed}
  I.~Zahed and G.E.~Brown,
  Phys.\ Rep.\  \textbf{142} (1986) 1

\bibitem{WirzbaA_MeiZahed}
  U.-G.~Mei{\ss}ner and I.~Zahed,
  Adv.\ Nucl.\ Phys.\  {\bf 17} (1986) 143

\bibitem{WirzbaA_Ulug2001}
  U.T.~Yakhshiev, M.M.~Musakhanov, A.M.~Rakhimov, U.-G.~Mei{\ss}ner, and A.~Wirzba, 
   Nucl. Phys.   \textbf{A700} (2002) 403

\bibitem{WirzbaA_Ulug2002}
  U.T.~Yakhshiev, U.-G.~Mei{\ss}ner and A.~Wirzba, 
   Eur. Phys.  J.  \textbf{A16} (2003) 569

\bibitem{WirzbaA_Ulug2004}
  U.T.~Yakhshiev, U.-G.~Mei{\ss}ner, A.~Wirzba, A.M.~Rakhimov, and M.M.~Musakhanov, 
   Phys.  Rev.  \textbf{C71} (2005) 034007

\bibitem{WirzbaA_Rathske}
  E.~Rathske,
  Z.\ Phys.\ A \textbf{331}  (1988) 499


\bibitem{WirzbaA_us1}
  U.-G.~Mei{\ss}ner, A.M.~Rakhimov, A.~Wirzba, and U.T.~Yakhshiev, 
   Eur. Phys. J.  \textbf{A31} (2007) 357


\bibitem{WirzbaA_us2}
  U.-G.~Mei{\ss}ner, A.M.~Rakhimov, A.~Wirzba, and U.T.~Yakhshiev, 
   Eur.  Phys.  J.   \textbf{A32}  (2007) 299

\bibitem{WirzbaA_us3}
  U.-G.~Mei{\ss}ner, A.M.~Rakhimov, A.~Wirzba, and U.T.~Yakhshiev, 
   Eur.  Phys.  J.  \textbf {A36}  (2008) 37

\bibitem{WirzbaA_Gasser}
  J.~Gasser and H.~Leutwyler,
  Phys.\ Rept.\  {\bf 87}  (1982) 77

\bibitem{WirzbaA_Thomas51}
  R.G.~Thomas, Phys.\ Rev.  \ {\bf 81}  (1951) 148

\bibitem{WirzbaA_Ehrman51}
  J.B.~Ehrman, Phys.\ Rev.  \ {\bf 81}  (1951) 412

\bibitem{WirzbaA_Bulgac96}
  A.~Bulgac, V.R.~Shaginyan, Nucl.\ Phys.\ A {\bf 601} (1996) 103

\bibitem{WirzbaA_Fayans98} 
  S.A.~Fayans,
  JETP Lett.\  {\bf 68} (1998) 169

\bibitem{WirzbaA_Bulgac99} 
 A.~Bulgac, V.R.~Shaginyan, Eur.\ Phys.\ J. A {\bf 5} (1999) 247

\bibitem{WirzbaA_Fayans01}
  S.A.~Fayans and D.~Zawischa,
  Int.\ J.\ Mod.\ Phys.\  B {\bf 15}  (2001) 1684

\bibitem{WirzbaA_Brown98}  B.A.~Brown,
  Phys.\ Rev.\  C {\bf 58} (1998) 220

\bibitem{WirzbaA_Brown00} B.A.~Brown,
 RIKEN Rev. {\bf 26} (2000) 53


\end{thebibliography}
\end{document}